# Epitaxial germanidation of full-Heusler $Co_2FeGe$ alloy thin films formed by rapid thermal annealing

Yota Takamura[1,2,a)], Takuya Sakurai[1,2], Ryosho Nakane[3], Yusuke Shuto[2], and Satoshi Sugahara[1,2,b)]

[1] *Department of Electronics and Applied Physics, Tokyo Institute of Technology, 4259 Nagatsuta-cho, Midori-ku, Yokohama 226-8503, Japan*

[2] *Imaging Science and Engineering Laboratory, Tokyo Institute of Technology, 4259, Nagatsuta-cho, Midori-ku, Yokohama 226-8503, Japan*

[3] *Department of Electrical Engineering and Information Systems, The University of Tokyo, 7-3-1 Hongo, Bunkyo-ku, Tokyo 113-8656 Tokyo, Japan*

___________________________________________________________________

a) Electronic mail: yota@isl.titech.ac.jp

b) Electronic mail: sugahara@isl.titech.ac.jp

**ABSTRACT**

The authors demonstrated that a full-Heusler Co$_2$FeGe (CFG) alloy thin film was epitaxially grown by rapid-thermal-annealing-induced germanidation of an Fe/Co/pseudo-Ge(001)-on-insulator (GOI) multilayer formed on a Si-on-insulator (SOI) substrate. X-ray diffraction (XRD) measurements with the out-of-plane and in-plane configurations revealed that the CFG film was epitaxially grown along the [001] direction with the in-plane epitaxial relation of CFG[100]∥GOI[100], although the film slightly contained a texture component. The strong (111) and (200) superlattice diffraction intensities indicated that the CFG film had a high degree of order for the *L*2$_1$ structure. Cross-sectional high-resolution transmission electron microscopy images of the film implied that the film had the dominant epitaxial and slight texture components, which was consistent with the XRD measurements. The epitaxial component was grown directly on the BOX layer of the SOI substrate without the formation of any interfacial layer.

**BODY**

Ge-channel metal-oxide-semiconductor field-effect transistors (MOSFETs)[1,2] are a very attractive candidate for next-generation high-performance MOSFETs owing to their high current drivability superior to Si-channel MOSFETs, which is the consequence of the high electron and hole mobilities of Ge. Recently proposed spin MOSFETs[3,4] using ferromagnetic electrodes for the source/drain (S/D) are a promising functional transistor with the variable transconductance that is controlled by the magnetization configuration of the ferromagnetic S/D. The spin MOSFETs are expected as a key device for a new class of highly functional logic circuit, e.g., nonvolatile power-gating systems.[5,6] Therefore, Ge-channel spin MOSFETs[3,4,7] would have a great impact on such highly functional logic circuits adaptable to the advanced Ge-channel MOSFET technology.

Half-metallic ferromagnets (HMFs)[8] that have a spin polarization of 100 % at the Fermi energy are the most promising ferromagnetic material for the S/D of spin MOSFETs. In particular, Co-based full-Heusler alloys, such as $Co_2FeSi$ (CFS)[7,9-13], $Co_2FeGe$ (CFG)[7,14,15], $Co_2MnSi$[16,17], and $Co_2FeSi_{1-x}Al_x$[18,19], are attractive, since they show a half-metallic band structure even at temperatures greater than room temperature.[20,21] In order to realize Ge-channel spin MOSFETs, a formation technique of the HMF S/D suitable for the complementary metal-oxide-semiconductor (CMOS) technology platform should be established. In advanced metal S/D CMOS devices, the S/D are comprised of a metal silicide that is formed by silicidation of transition metals or rare earth metals deposited on the S/D region of a Si substrate.[22] In general, rapid thermal annealing (RTA) is commonly used for the activation of the silicidation reaction. In a similar manner, metal germanides formed by RTA-induced germanidation can be used for Ge channel MOSFETs.[23,24] Recently we reported that Si-containing full-Heusler CFS alloy thin films were successfully formed by RTA-induced silicidation of an Fe/Co/Si-on-insulator (SOI) multilayer.[7,10-12] The stoichiometry of the RTA-formed CFS thin films was controlled by adjusting the film thicknesses of each layer.[10,11] The crystallographic features of the RTA-formed CFS films were (110)-oriented texture of $L2_1$-structure grains with excellent crystallinity.[10-12] In particular, the film formed at 800 °C exhibited a high degree of $B2$-order of 97 % and also a high degree of $L2_1$-order of 83 %.[11] Moreover, the CFS film contained a very small amount of the $DO3$ disorder.[12] In an analogous fashion to Si-containing full-Heusler alloys, the RTA technique[10,25] can be applied to Ge-containing full-Heusler alloys: Full-Heusler CFG alloy thin films were successfully formed by RTA-induced germanidation of an Fe/Co/Ge-on-insulator (GOI) multilayer.[7]

In this study, we demonstrated that an RTA-formed CFG film was epitaxially grown by the germanidation reaction of an Fe/Co/GOI multilayer. It is known that particular silicides, such as $NiSi_2$ whose lattice constants are very close to that of Si, are epitaxially grown during the silicidation.[26] The lattice mismatch between CFG(001) and GOI(001) is also small (~1 %)[7,14], and thus epitaxial germanidation growth is also expected for RTA-formed CFG films.

For the formation of CFG thin films, pseudo-GOI substrates[7,27] consisting of an epitaxially grown Ge(001) film on a thinned SOI(001) layer of an SOI substrate were prepared: First, a commercially available wafer-bonding SOI substrate was thermally oxidized until the SOI layer was thinned to 2 nm. After the etching of the resulting oxide layer, a 40-nm-thick Ge(001) film was epitaxially grown by low-temperature molecular beam epitaxy (LT-MBE) technique on the thinned SOI layer. The epitaxial Ge film exhibited a very flat surface without Stranski-Krastanov (S-K) mode growth islands.[28] Hereafter, the epitaxial Ge layer is referred to as a GOI layer for simplicity. Co and Fe films were deposited on the GOI layer in an ultrahigh vacuum by electron beam evaporation method. Film thicknesses of Co and Fe layers were set to 44 nm and 21 nm, respectively, which were determined so that the stoichiometric composition of $Co_2FeGe$ was achieved. Subsequently, the germanidation was performed by RTA in $N_2$ atmosphere at an RTA temperature ($T_A$) of 750 ºC for an RTA time ($t_A$) of 4 min. The atomic concentration of Co, Fe, Ge, and Si in the CFG film was measured by secondary ion mass spectroscopy (SIMS) (that was calibrated by Rutherford backscattering spectrometry (RBS) and particle induced x-ray emission (PIXE)). Co, Fe, and Ge atoms were homogeneously distributed in the film and their concentrations in the CFG film were 57 %, 23 %, and 20 %, respectively, which was close to the stoichiometric composition. The Si atoms in the ultrathin SOI layer were distributed uniformly in the film, and the concentration was less than 1 % in atomic percentage.

Structural properties of the CFG film were analysed by means of x-ray diffraction (XRD) with both out-of-plane and in-plane configuration measurements using a Cu $K\alpha$ source. The definitions of the multiaxis goniometer geometries of the out-of-plane and in-plane configurations are shown in Figs. 1(a) and (b), respectively. Figure 2 shows the $\omega$–$2\theta$ scan pattern of the RTA-formed CFG film, measured with the out-of-plane configuration (where the incident x-ray was set to be parallel to the Si(100) plane, which defined zero for the in-plane rotation angle $\phi$ and for the tilt angle $\psi$). CFG(002), CFG(022), and CFG(004) diffraction peaks were clearly observed, indicating that the CFG thin film had two orientational components of the (001) and (011) faces. The estimated intensity ratio $I_{004}/I_{002}$ between the CFG(004) and CFG(022) diffraction peaks of the film was ~64 that was much higher than that (0.1) of powder CFG[29]. Therefore, the (001)-oriented component was dominant. Furthermore, the observed CFG(002) superlattice diffraction indicated that the CFG film had at least the *B*2-ordered structure.

The detailed crystallographic features of the RTA-formed CFG thin film were investigated by in-plane XRD measurements. In the in-plane configuration, $2\theta\chi$ is an angle between the incident x-ray and the detector, and $\phi$ is the in-plane rotation angle of the stage, as shown in Fig. 1(b). The sample was set so that the incident x-ray was parallel to the Si(100) plane of the sample for $\phi = 0°$. Note that $\phi$ is equal to $\theta\chi$ during $\phi$–$2\theta\chi$ scans. When the $\phi$–$2\theta\chi$ scan started from the initial $\phi$ angle ($\phi_{ini}$) of 0°, the $\phi$–$2\theta\chi$ scan pattern of the film showed strong CFG(200) and CFG(400) diffraction peaks, and an week CFG(022) diffraction peak, as shown in Fig. 3(a). On the other hand, a strong CFG(220) diffraction peak and a week CFG(400) diffraction peak were observed, when the $\phi$–$2\theta\chi$ scan started from $\phi_{ini} = 45°$, as shown in Fig. 3(b). Solid and thin curves in Fig. 3(c) show the $\phi$ scan patterns for the CFG(400) diffraction in Fig. 3(a) and CFG(220) diffraction in Fig. 3(b), respectively, in which they were measured with $2\theta\chi$ fixed to the corresponding Bragg angles for each diffraction ($2\theta\chi_B = 64.9°$ for the CFG(400) diffraction and 44.6° for the CFG(220) diffraction). The horizontal axis in Fig. 3(c) is $\phi - \theta\chi_B$ that represents the relative in-plane rotation angle measured from the Si(100) plane. The solid curve in Fig. 3(c) shows the ($\bar{4}$00), (0$\bar{4}$0), (400), and (040) diffraction peaks of CFG: These were periodically separated each other with the angle difference of 90°, showing fourfold symmetry in the sample plane. The ($\bar{2}$20), ($\bar{2}\bar{2}$0), (2$\bar{2}$0), and (220) diffraction also exhibited fourfold symmetry, as shown by the thin curve in Fig. 3(c). The (400) and (220) diffraction peaks in the $\phi$ scan patterns were separated from 45°, which corresponded to the crystal structure of CFG. These results indicated that the dominant (001) component shown in Fig. 2 was epitaxially grown with the in-plane epitaxial relation of CFG[100]‖GOI[100] (that is parallel to Si[100] of the SOI substrate). On the other hand, the weak CFG(220) and CFG(400) diffraction in Figs. 3(a) and (b), respectively, showed the $\phi$-independent constant signals in the $\phi$–scan patterns that appeared in Fig. 3(c). Although these signals look like background, they include a contribution from diffraction). Taking the results shown in Figs. 2 and 3 into account, we concluded that the (011)-orientational component shown in Fig. 2 was texture.

The CFG(111) diffraction of the CFG film was clearly observed with the goniometer configuration of $\psi = 54.6°$ and $\phi = 45°$, indicating the $L2_1$ structure. The $\phi$ scan pattern for the CFG(111) diffraction shown in Fig. 3(d) exhibited fourfold symmetry, which also supported the epitaxial germanidation of (001)-oriented component of the CFG film.

The order parameters for the CFG film were evaluated using the extended Webster model[11,29]. The degree of long-range order for the $L2_1$ structure and that for the $B2$ structure can be evaluated using indices $S_{L21}$ and $S_{B2}$.[11,29] These were deduced using the superlattice diffraction intensity ratios of CFG(111) to CFG(202), measured with the out-of-plane configuration, and of CFG(200) to CFG(400), measured with the in-plane configuration. $S_{L21}$ and $S_{B2}$ showed high values, e.g., $S_{L21} = 83$ %, and $S_{B2} = 85$ %. Note that the estimation of these values requires reference data[30,31] for fully $L2_1$-ordered CFG, and thus the estimated values depend on the reference data.

Figure 4(a) shows the cross-sectional high-resolution transmission electron microscopy (HRTEM) image of the CFG film. The overall image showed relatively uniform contrast, although it contained defects and boundaries, implying that the CFG film was mostly epitaxial. The inhomogeneous parts would be the texture component described above. The sample surface was very flat, which may reflect the epitaxial growth of the film. Figure 4(b) shows the magnified view of the HRTEM image. The epitaxial component was grown directly on the BOX layer of the SOI substrate without the formation of any interfacial layer. The bright and dark linear-shape contrasts near the interface would result from a tiny interface roughness or roll, in which the depth-directional interface roughness/roll can be transparently seen

In summary, we demonstrated that a full-Heusler CFG alloy thin film was epitaxially grown by RTA-induced germanidation of an Fe/Co/pseudo-GOI(001) multilayer formed on an SOI substrate. XRD measurements with the out-of-plane and in-plane configurations revealed that the CFG film was epitaxially grown along the [001] direction with the in-plane epitaxial relation of CFG[100]∥GOI[100], although the film slightly contained a texture component. The strong (111) and (200) superlattice diffraction intensities indicated that the CFG film had high a degree of order of the $L2_1$ structure. Cross-sectional HRTEM images of the film implied that the film had the dominant epitaxial and slight texture components, which was consistent with the XRD measurements. The epitaxial component was grown directly on the BOX layer of the SOI substrate without the formation of any interfacial layer.

The authors would like to thank Prof. H. Munekata, Tokyo Institute of Technology, Profs. M. Tanaka and S. Takagi, The University of Tokyo.

**Figure captions**

Fig. 1: Definitions of multiaxis goniometer geometries of (a) out-of-plane and (b) in-plane configurations.

Fig. 2 XRD $\omega$–$2\theta$ scan pattern of the RTA-formed CFG film measured with the out-of-plane configuration, where the incident x-ray was set to be parallel to the Si(100) plane, which defined zero for the in-plane rotation angle $\phi$ and for the tilt angle $\psi$.

Fig. 3 In-plane XRD patterns for (a) $\phi$–$2\theta\chi$ scan when it started from $\phi_{ini} = 0°$, (b) $\phi$–$2\theta\chi$ scan when it started from $\phi_{ini} = 45°$, and (c) $\phi$ scan for CFG(400) and CFG(220). The solid and thin curves show diffraction for the (400) and (220), respectively. (d) Out-of-plane XRD pattern for $\phi$ scan for CFG(111).

Fig. 4 Cross-sectional HRTEM images of the RTA-formed CFG thin film.

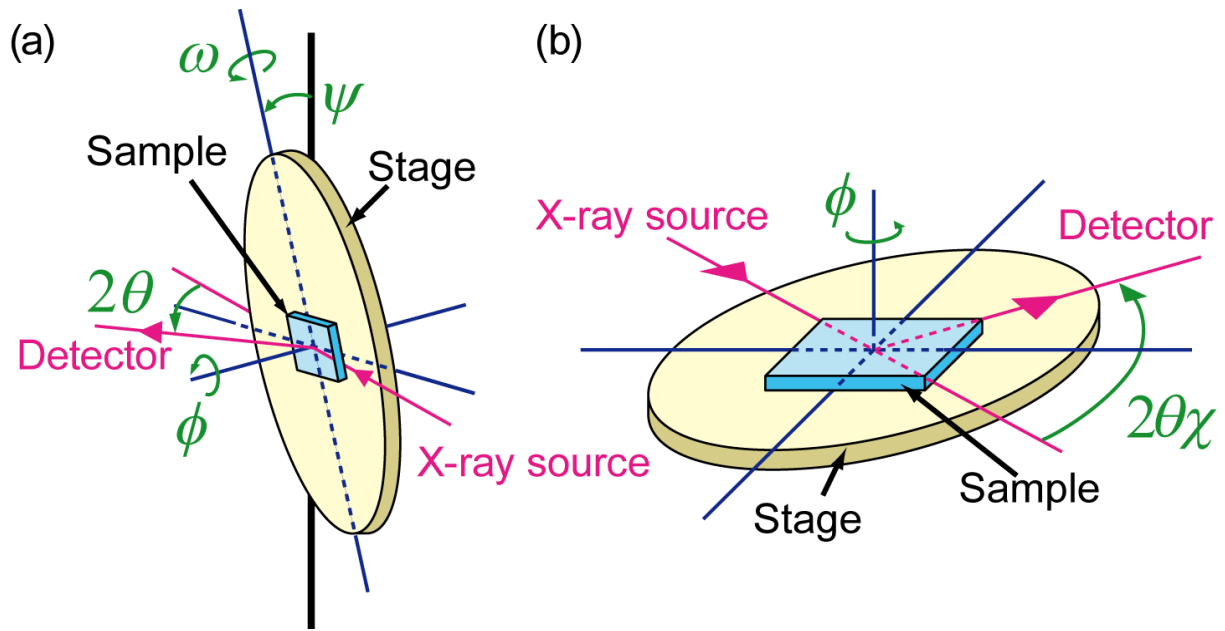

Figure 1 takamura *et al*.

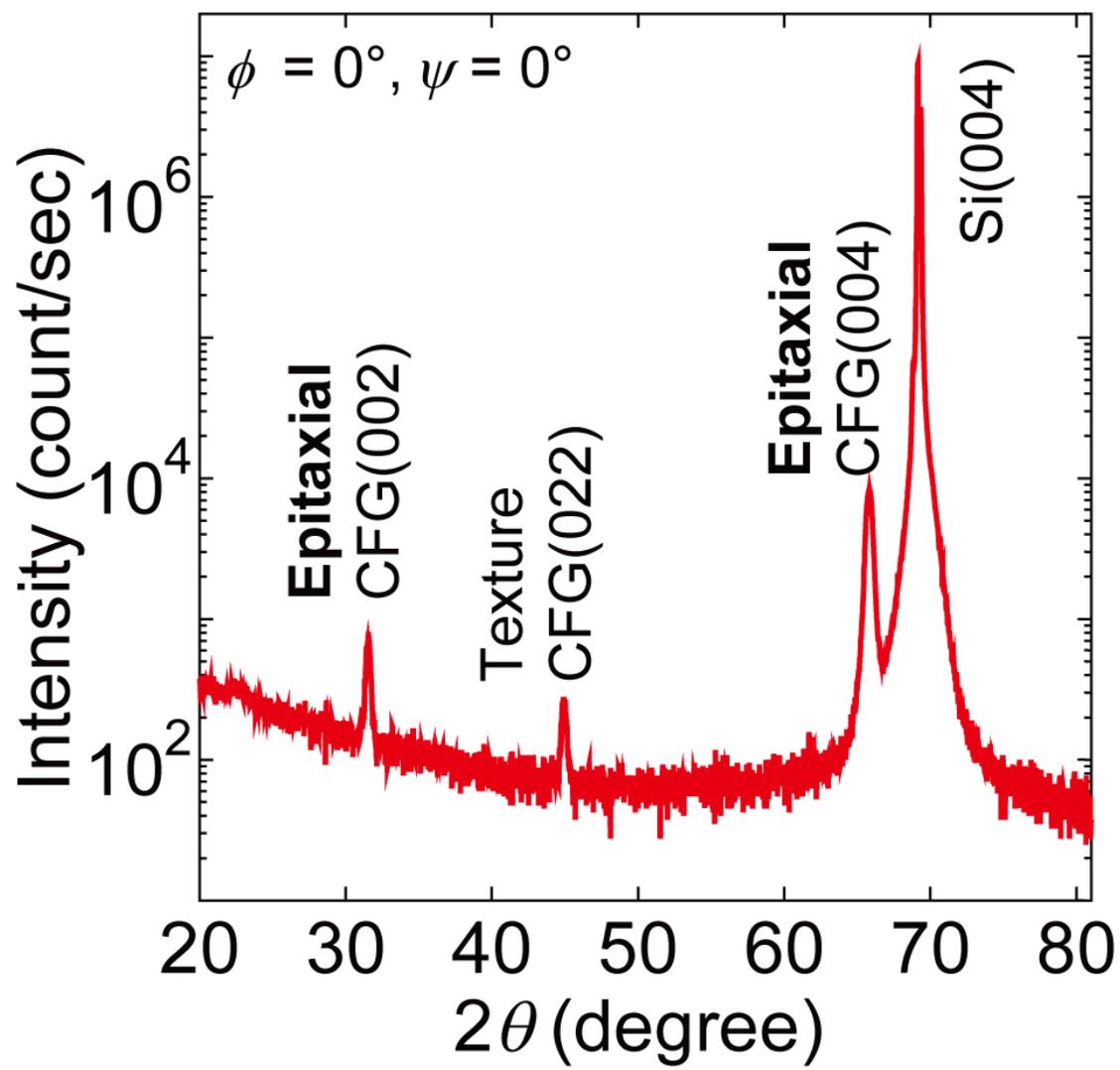

Figure 2 takamura *et al*.

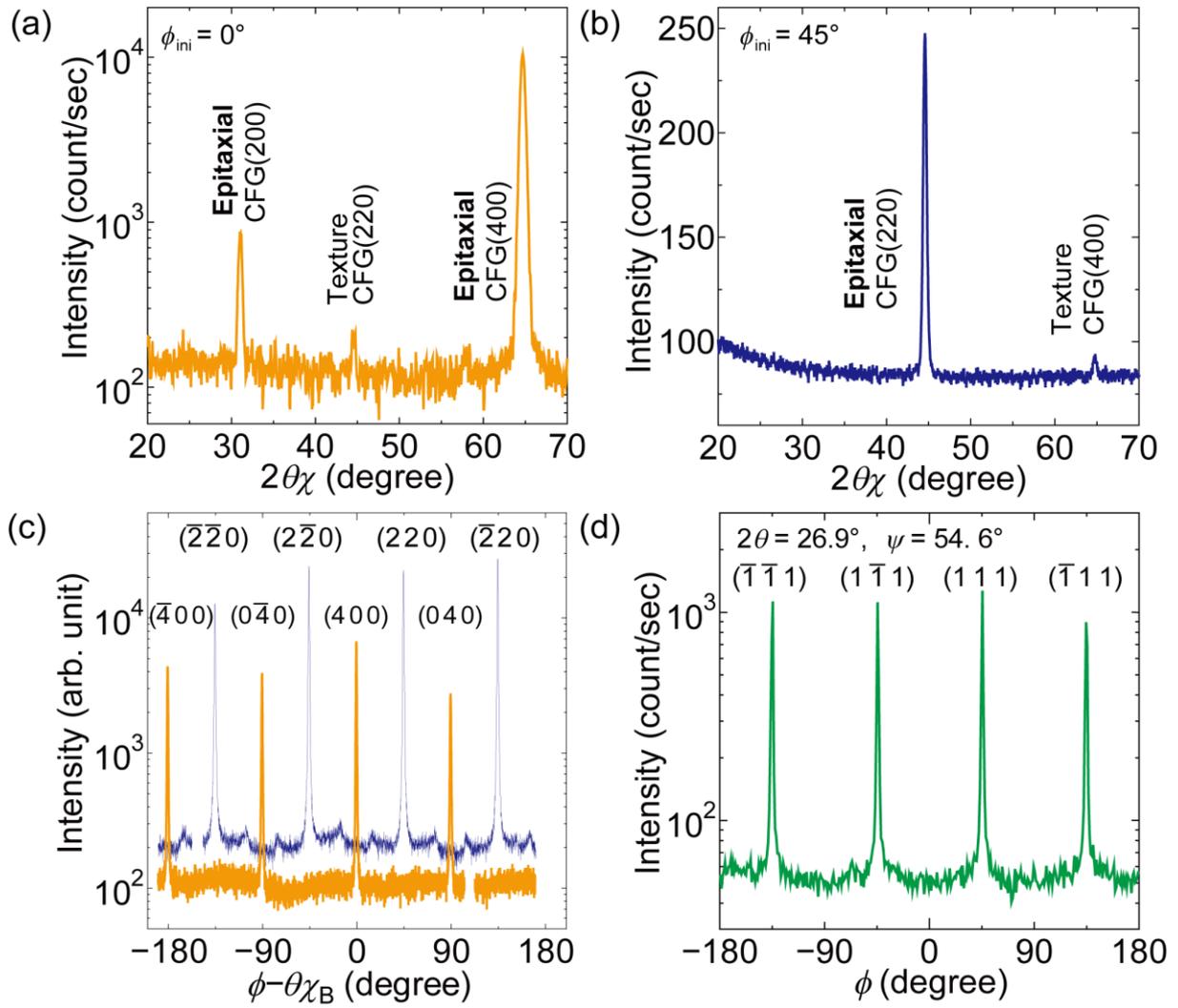

Figure 3 takamura *et al*.

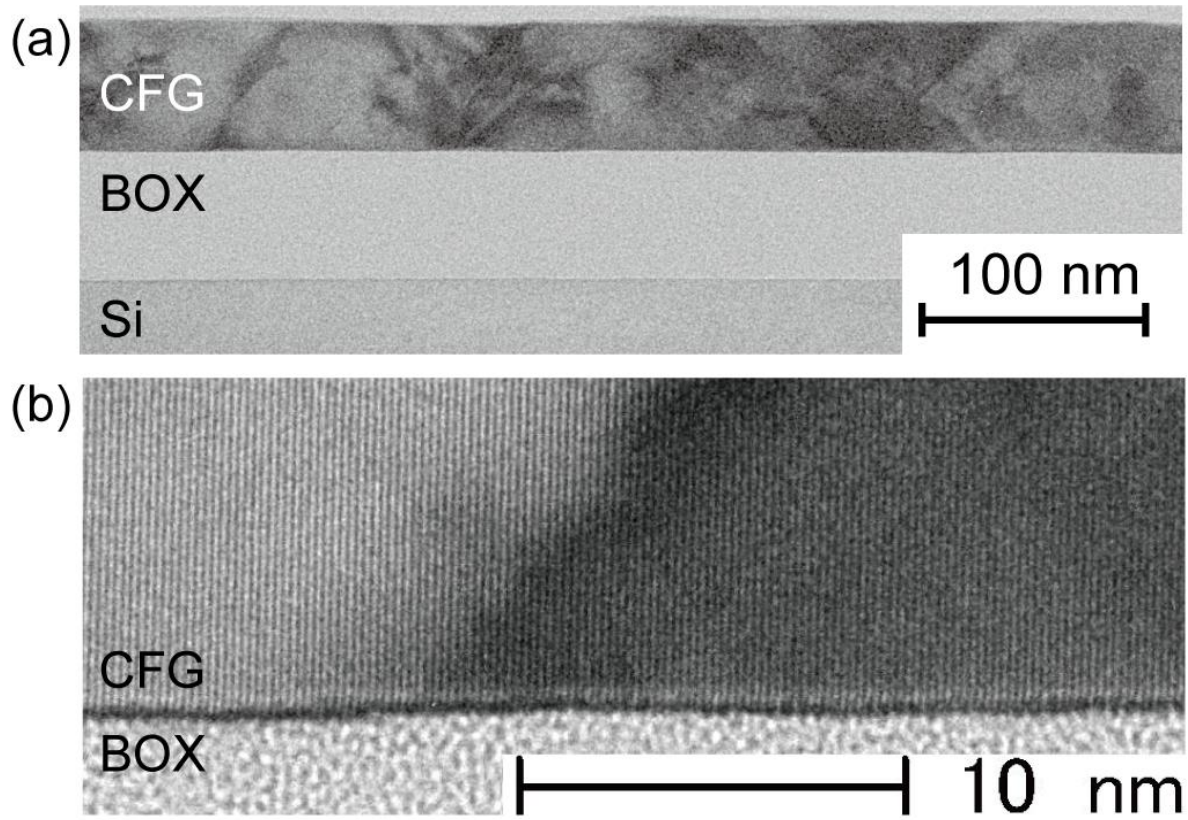

Figure 4 takamura *et al*.